\documentclass[aps,showpacs,eqsecnum]{revtex4}

\usepackage{graphicx}
\usepackage{amsmath}
\usepackage{amssymb}
\usepackage[footnotesize,sc]{caption}

\usepackage{color}

\def\b{\beta}
\def\e{\varepsilon}
\def\d{\delta}

\def\m{\mu}
\def\t{\tau}
\def\n{\nu}
\def\o{\omega}
\def\s{\sigma}
\def\G{\Gamma}

\def\D{\Delta}
\def\O{\Omega}

\def\ra{\rightarrow}

\def\pd{\partial}
\def\bk{{\bf k}}
\def\bp{{\bf p}}
\def\bq{{\bf q}}
\def\br{{\bf r}}
\def\bQ{{\bf Q}}

\def\be{\begin{equation}}\def\ee{\end{equation}}
\def\bea{\begin{eqnarray}}\def\eea{\end{eqnarray}}

\def\pref#1{(\ref{#1})}
\newcount\bozza \bozza=0
\ifnum\bozza=1
\newdimen\shift \shift=-2truecm
\def\lb#1{%
{\label{#1}\rlap{\kern\shift{$\scriptstyle#1$}}}}
\else\def\lb#1{\label{#1}} \fi

\begin{document}

%\preprint{HEP/123-qed}

\title{Optical-conductivity sum rule in cuprates
and unconventional charge density waves:  a short review}

\author{L.~Benfatto$^1$}
%\email{lara.benfatto@roma1.infn.it}

\author{S.G.~Sharapov$^{2}$}
%\email{sharapov@bitp.kiev.ua}
%\homepage{http://www.geocities.com/sergei_sharapov/}

\affiliation{
$^1$SMC-INFM-CNR and Department of Physics, University of Rome ``La
  Sapienza'',\\ Piazzale Aldo Moro 5, 00185, Rome, Italy\\
$^2$Department of Physics and Astronomy, McMaster University,
        Hamilton, Ontario, Canada, L8S 4M1}

\date{\today }

\begin{abstract}
We begin with an overview of the experimental results for the
temperature and doping dependences of the optical-conductivity
spectral weight in cuprate superconductors across the whole phase
diagram. Then we discuss recent attempts to explain the observed
behavior of the spectral weight using reduced and full models with
unconventional  $d_{x^2-y^2}$ charge-density waves.
\end{abstract}


%71.10.-w        Theories and models of many-electron systems
%74.25.Gz        Optical properties
%72.15.-v        Electronic conduction in metals and alloys
%74.72.-h        High-Tc compounds

%\keywords{}

\maketitle

\section{Preliminaries}

The optical-conductivity sum rule plays an important role in the
understanding of the physics of cuprate superconductors. The behavior of
the spectral weight in the pseudogap phase and when the system undergoes a
superconducting transition has been the object of the experimental
investigation in the last few years
\cite{Molegraaf:2002:Science,Santander:2002:PRL,Santander:2004,Homes:2004:PRB,Boris:2004:Science,Ortolani:2005:PRL,Deutscher:2005,Carbone:2005}.
Considering that the analysis of the experimental data and the extraction
of the spectral weight involves a lot of theoretical assumptions, so that
even opposite conclusions can be drawn based on the same data
\cite{Boris:2004:Science,Santander:2005,Kuzmenko:2005}, we decided to
describe the current experimental situation in the introductory
Sec.~\ref{sec:general} using the theoretical language from the very
beginning, to avoid any possible confusion. We find that a clear perception
of these assumptions is crucial for the correct interpretation of the
experiments.

In Sec.~\ref{sec:reduced_model} we briefly summarize the recent history of
unconventional density waves, and introduce the reduced $d$-density wave
(DDW) model, which can be regarded as the charge density wave equivalent of
the reduced BCS model. Then based on
Refs.~\cite{Benfatto:2004:EPJ,Benfatto:2005:PRB} we discuss the optical
conductivity and the sum rule for the reduced model, as an attempt to
interpret some of the recent data on the sum rule of cuprates. In
Sec.~\ref{sec:full-model} we discuss the limitations of the reduced model,
as already pointed out clearly in Ref.~\cite{Benfatto:2005:PRB}, and
consider the recent results obtained for a more generic microscopic
Hamiltonian by Aristov and Zeyher \cite{Aristov:2005}. Some of the
remaining questions and the relationship to other systems are discussed in
Sec.~\ref{sec:questions}.

\section{General notations and overview of experiments}
\lb{sec:general}

%\section{The general notation}

\subsection{Optical conductivity and sum rule for it}
\label{sec:general-sigma-sum}

The optical-conductivity sum rule is intimately related to the
general properties of the current-current response function
\cite{Millis:book,Mahan}. Since we are interested in describing
lattice systems, let us consider a general case of the electronic
system described by the Hamiltonian:
\be \lb{Hamiltonian} H = H_0+H_{int},\qquad H_0= -\sum_{ij}t_{ij}
c_{i\sigma}^{\dagger}c_{j\sigma}-\mu\sum_{i}
c_{i\sigma}^{\dagger}c_{i\sigma}, \ee
where the field operator $c^{\dagger}_{i\sigma}$ creates an electron of
spin $\sigma$ at the site $\mathbf{r}_i$, $t_{ij}$ is the hopping
parameter, $\mu$ is the chemical potential and $H_{int}$ is the interaction
term. For the kinetic part $H_0$ a proper description of the cuprates is
obtained by restricting the sum to next-neighboring sites on a squared
lattice that results in a band dispersion $\varepsilon(\mathbf{k}) = - 2 t
(\cos k_x a + \cos k_y a)+4t' \cos k_x a\cos k_y a$, where $a$ is the
lattice constant. However, to avoid unnecessary complications we will also
refer below to the simplest case $t'=0$. Throughout the paper units $\hbar
= k_B =c =1$ are chosen.

Given the model \pref{Hamiltonian}, the definitions of the particle
current ${\bf j}(\br_i)$ and the so-called diamagnetic tensor
$\t_{ii}(\br_i)$ follow from the first- and second-order derivatives
of the Hamiltonian $H({\bf A})$ with respect to the external vector
potential ${\bf A}$  \cite{Scalapino:1993:PRB}:
\be
\lb{j-tau.def} H(A_i)\approx H(0)  - \sum_{j} \left[ e
A_i(\mathbf{r_j}) j_i(\mathbf{r_j}) - \frac{e^2}{2}
A_i^2(\mathbf{r_j}) \tau_{ii}(\mathbf{r_j}) \right],
\ee
so that the total electric current density is expressed as
\be
J_i (\mathbf{r})= -
\delta H/\delta A_i(\mathbf{r}) = e j_i(\mathbf{r}) - e^2
\tau_{ii}(\mathbf{r}) A_i(\mathbf{r}),
\ee
By using the generalized notation $j_{\mu}=(j_i,j_0)$ where the
index $\mu=(i,0 )$ with $i=1,2$ indicates spatial and time
components respectively, and $j_0$ is the particle density, one can
evaluate $\langle J_\m (q) \rangle$ within  linear response theory
\cite{Mahan,Scalapino:1993:PRB,Schrieffer} and obtain $J_{\mu}(q)=e^2
K_{\mu\nu}(q)A_{\nu}(q)$, where the electromagnetic kernel
$K_{\m\n}$ is defined as:
\be
\lb{em-kernel} K_{\m\n}(\bq,
i\O_m) =-\tau_{\m\m}\d_{\m\n}(1-\d_{\n0})+\Pi_{\m\n}(\bq, i\O_m).
\ee
Here $\Pi_{\mu\nu}(\bq, i \O_m)$ is the current-current correlation function
\be
\lb{cur-cur}
\Pi_{\mu\nu}(\bq, i\O_m)= \frac{1}{N}\int_0^\b
d\t e^{i\O_m\t}\langle T_\t j_\m(\bq, \t )j_\n( -\bq, 0 )
\rangle,
\ee
$\tau$ is the imaginary time, $\b = 1/T$, $\O_m = 2 \pi m T$ is
the bosonic Matsubara frequency and $N$ is the number of unit cells.
Finally, using that $\mathbf{A}(\o) = \mathbf{E}(\o)/i(\o+i0)$,
where $\mathbf{E}$ is the electric field, one arrives at the famous
Kubo formula
\be
\lb{conductivity.def} \s(\o)=-ie^2\frac{K_{i
i}(\bq=0,\o)}{V(\o+i0)}= ie^2\frac{<\tau_{i i}>-\Pi_{i
i}(\bq=0,\o)}{V(\o+i0)},
\ee
where $V$ is the unit-cell volume and the standard analytic
continuation $i \Omega_m \to \omega +i 0$ was made. To avoid
confusion,  we will indicate the imaginary bosonic frequencies with
$i\O_m$, the imaginary fermionic frequencies with $i\o_n = i \pi
(2n+1) T$ and the real frequencies with $\o$. Since an isotropic
system is considered we can omit the index $i$ as done in the LHS of
Eq.~\pref{conductivity.def} and in what follows.

To extract the real and imaginary parts of the optical conductivity
defined in Eq.\ \pref{conductivity.def} we should consider the
effect of disorder. Indeed, for the (ideal) case of a system without
disorder by taking the real part of \pref{conductivity.def} one
obtains $\mbox{Re} \s(\o)=(\pi e^2/V) \d(\o)[<\t>-\mbox{Re}
\Pi(\mathbf{0},\o)]+ (e^2/V) \mbox{Im}\Pi(\mathbf{0},\o)/\o$. The
presence of a delta function, which means an ideal conductivity of
the system, is an artifact of the assumption that no impurities are
present. As soon as  disorder is present one expects that in the
normal state the following identity holds:
\be
\lb{nodis}
\mbox{Re} \Pi(\bq=0,\o\ra0)=<\t>,
\ee
so that only the regular part of $\s(\o)$ survives, defined as usual as:
\be
\lb{Re.sigma}
\mbox{Re} \s(\o)= \frac{e^2}{V} \frac{\mbox{Im} \Pi(\bq = \mathbf{0},\o)}{\o}.
\ee
Thus, using the Kramers-Kronig (KK) relations for $\Pi(\bq =0,\o)$
one can derive the well-know sum rule:
\be
\lb{common.rule}
W(T)=\int_{-\infty}^{\infty}
\mbox{Re} \sigma(\omega) d \omega = \frac{e^2}{V} \int_{-\infty}^{\infty}
\frac{\mbox{Im} \Pi(\bq = \mathbf{0},\omega)}{\omega}
d \omega = \frac{\pi e^2}{V}
\mbox{Re}\Pi(\bq = \mathbf{0},\o=0)=\frac{\pi e^2}{V} <\tau>.
\ee
Observe that Eq.~\pref{nodis} and \pref{common.rule} require that
for the system in the presence of disorder, the {\em dynamic}
$(\bq=0,\o\ra 0)$ and {\em static} $(\o=0,\bq\ra 0)$ limits of
$\mbox{Re}\Pi(\bq,\o)$ commute. This is indeed the case for a
disordered system, but not for a clean one, where usually $\mbox{Re}
\Pi({\bf 0},\o)$ vanishes as $\o\ra 0$ \cite{Schrieffer,
Scalapino:1993:PRB}.

As one can see from Eq.~\pref{common.rule} the sum rule is directly
defined by the diamagnetic tensor, which in turn depends, according
to Eq.~\pref{j-tau.def}, on the way how the vector potential ${\bf
A}$ enters the Hamiltonian of the system. When a continuum model is
considered instead of Eq.~\pref{Hamiltonian}, the kinetic term is
expressed as $\int (-\nabla)^2/2m$ and ${\bf A}$ is inserted using
the minimal coupling prescription $- i \nabla \to - i \nabla - e
\mathbf{A} $. For lattice systems the equivalent of the minimal
coupling prescription is the so-called {\em Peierls ansatz\/}
\cite{Jaklic:2000:AP,Millis:book,Scalapino:1993:PRB}, which
corresponds to inserting the gauge field ${\bf A}$ in
Eq.~\pref{Hamiltonian} by means of the substitution $c_i\ra c_i e^{-
i e \int {\mathbf A}\cdot d {\mathbf r}}$. In this case, it is clear
that when the interaction term of the Hamiltonian is a
density-density interaction, as in Eq.~\pref{Hubbard} below, only
the kinetic hopping term is modified, while the interaction term is
gauge invariant (GI).  As a result, the current/density operator and
the diamagnetic tensor can be expressed (for small ${\bf q}$) as:
\bea \lb{defj} j_\m(\bq,t)&=&\frac{1}{N}\sum_{\mathbf{k}, \sigma}
v_\mu(\bk) c^{\dagger}_{\mathbf{k}- \mathbf{q}/2\s}
c_{\mathbf{k}+\mathbf{q}/2\s}, \\\lb{deft}
\langle\tau_{ii}\rangle&=&\frac{1}{N}\sum_{\bk,\s} \frac{\partial^2
\e_\bk }{\partial k_i^2} n_{\bk,\sigma}, \eea
where
\be \lb{vbare} v_\mu(\bk)=(v_\bk^F,1), \ee
and $(v_{\bk}^F)_i=\pd \e_\bk/\pd k_i$ is the Fermi velocity. As a
consequence, we obtain the restricted optical-conductivity sum rule:
\be
\lb{weight.band}
W(\o_M,T)=\int_{-\omega_M}^{\omega_M} \mbox{Re} \,
\sigma_{ii}(\omega,T)d \omega
\equiv W(T) = \frac{\pi e^2}{V} \langle
\tau_{ii} \rangle=
\frac{\pi e^2}{V N} \sum_{\mathbf{k},\sigma}
\frac{\partial^2 \e_\bk }{\partial k_i^2} n_{\bk,\sigma}
= - \frac{\pi e^2a^2}{V} \frac{\langle K \rangle}{n_d} ,
\ee
where the last equality, relating the spectral weight to the average
value of the kinetic energy $K$, only holds for a nearest-neighbor
tight-binding model ($t'=0$), and $n_d$ is the dimensionality of the
system. Observe also that here $K$ includes the sum over two spin
components, so that the factor $1/n_d$ comes from the fact that
$\sum_\bk \pd^2\e_\bk/\pd k_i^2=\sum_\bk 2ta^2\cos k_i
a=-(a^2/n_d)\sum_\bk \e_\bk$. In writing Eq.~\pref{weight.band} we
introduced a cut-off $\o_M$ in the frequency integration
to make contact with the experiments, where only a limited range of
frequency is accessible. However, the equivalence we established
between Eq.~\pref{common.rule} and Eq.~\pref{weight.band} relies on
the fact that even though in deriving Eq.~\pref{common.rule} an
integration up to an infinite cut-off is formally required, an {\em
intrinsic} cut-off is provided by the energy $\o_M$ below which the
tight-binding description \pref{Hamiltonian} is valid. As we shall
see, the definition of such a cut-off in cuprates is a quite
delicate issue. Indeed, it is clear from Eq.~\pref{weight.band} that
the spectral weight depends both on temperature and interaction
trough the occupation number $n_{\bk,\s}$, and as a consequence it
is a powerful tool to test theoretical predictions coming from
different models.  This behavior should be contrasted to the case of
an electronic system described by a quadratic band dispersion
$\e_\bk = \bk^2/2m$. Indeed, in this case the tensor $\langle
\tau_{ij} \rangle$ reduces to $n \delta_{ij}/m$, where $n$ is the
total carrier density, so that Eq.~\pref{common.rule} reduces to the
so-called f-sum rule
\begin{equation}
\lb{full.rule}
\int_{-\infty}^{\infty} \mbox{Re} \, \sigma(\omega) d \omega
= \frac{\pi n e^2}{m},
\end{equation}
which is temperature and interaction independent. To make contact
between the restricted and full f-sum rule it can be useful to
define the frequency-dependent spectral weight:
\begin{equation}
\lb{weight}
W(\omega,T) = \int_{-\omega}^{\omega} \mbox{Re} \,
\sigma(\omega',T)d \omega'.
\end{equation}
In a lattice system one would expect that below $\o= \o_M$ only
transitions within the conduction band are included, while at higher
energies  interband processes play also a role,
\be
\s(\o)=\s_{intra}(\o)+\s_{inter}(\o),
\ee
such that $\s_{intra}(\o)=0$ at $\o>\o_M$ and $\s_{inter}(\o)=0$ at
$\o<\o_M$. As a consequence, the spectral weight $W(\o_M,T)$ only
measure $\s_{intra}(\o)$ and gives a measure of the interactions
acting on the system through the relation \pref{weight.band}. When
the integration frequency of Eq.~\pref{weight} is extended above
$\o_M$ one should evolve towards the f-sum rule \pref{full.rule},
and the inclusion of the processes described by $\s_{inter}(\o)$
should give a temperature independent spectral weight. Observe that
while many theoretical results have been provided in the literature
about the behavior of the restricted spectral weight
\pref{weight.band}, based on different interacting model as
Eq.~\pref{Hamiltonian}, no clear theoretical understanding exists
yet about the evolution of the spectral weight towards the f-sum
rule as $\o>\o_M$ \cite{Millis:book}.

\subsection{Conventional spectral-weight behavior}

The relation between the optical-conductivity spectral weight and
the diamagnetic tensor (or kinetic energy) introduced in the
previous section is particularly useful from the theoretical point
of view, because it allows one to derive the restricted sum rule
without doing an explicit computation of  the optical conductivity.
However, one can also be interested on the precise structure of
$\s(\o)$, and on the eventual transfer of spectral weight through a
phase transition. Before discussing the experimental data in the
cuprates it is worth showing some examples of the applicability of the
previous formulas in simple paradigmatic cases.

\begin{itemize}
\item{\bf Temperature dependence in the non-interacting nearest-neighbor
tight-binding system}

In this case, in Eq.\ \pref{weight.band} the occupation number is simply given
by the Fermi function, $n_{{\bf k} \sigma} =f(\xi_\bk)$, where
$\xi_\bk=\e_\bk-\mu$.  The main temperature dependence of the
spectral weight \pref{weight.band} comes from the temperature
smearing of the Fermi function, and can be easily evaluated using
the Sommerfeld expansion. In $n_d=2$ dimensions one obtains
\be \lb{weight.normal} \frac{W(T)}{(\pi e^2
a^2/V)}=-\frac{1}{N}\sum_\bk \e_\bk f(\xi_\bk)= -\int d \e N(\e)
\e f(\e-\mu)=\frac{W(0)}{(\pi e^2 a^2/V)}-c(\mu)T^2, \ee
where $c(\e)= (\pi^2/6)[\e N'(\e)+N(\e)]$, and $N(\e)$ and $N'(\e)$
are the density of states (per spin) of the tight-binding dispersion
and its derivative, respectively. For a flat density of states $N=1/2D$
where $D$ is the semi-bandwidth, so that:
\be
\lb{conventional}
\tilde W(T)=\tilde W(0)-\frac{\pi^2}{12D}T^2=\tilde W(0)-BT^2.
\ee
Observe that for a layered system $V=a^2d$, where $d$ is the
inter-plane distance, so that the quantity $\tilde W=W/(\pi e^2
a^2/V)$ has the dimension of an energy (see discussion Sec. II-C
below). The hopping $t$ of the two-dimensional band dispersion can
be related to the semi-bandwidth as $D=4t$. Analogously, one can
consider the tight-binding dispersion for $t'=0$ and estimate, at
small electron density, $N(\e)=1/4\pi t$. The slope $c=\pi/24t$ so
obtained is not far from the value of $c$ obtained around half
filling, see Ref.~\cite{Benfatto:2005:PRB}.

\item{\bf Cut-off definition in the Drude model}

In metals displaying a conventional Fermi-liquid behavior the
quasiparticles behave as free electrons with a renormalized mass $m^*$ and
a characteristic scattering time $\tau=1/\G$. In the Drude approximation the
real part of the optical conductivity reads:
\be
\lb{drude}
{\rm Re}\s_{Drude}(\o,T)=
\frac{ne^2\tau/m^*}{1+\o^2\t^2}=\frac{\G}{\pi}\frac{W(T)}{\G^2+\o^2},
\ee
where we made the dependence on the lattice spectral weight $W$
explicit. In Fig.~\ref{fig-drude} we report the integrated spectral
weight as a function of the frequency. As one can see, at frequencies
of the order of few times the inverse scattering time $\Gamma$,
$90\%$ of the total value $W(T)$ is recovered, showing that in this
case the cut-off frequency $\o_M$ can be identified as $\o_M\approx
6 \Gamma$ \cite{Homes:2004:PRB}. In a conventional metal one expects
that Eq.~\pref{drude} reproduces in good approximation the optical
conductivity, eventually with a $\Gamma$ slightly temperature
dependent. Moreover, the Fermi-liquid paradigm ensures that the
quasiparticles still have a Fermi surface, with an occupation number
$n_{\bk\s}$ given by the Fermi function. The only effect of
interactions is to renormalize the bandwidth, which can be for
example determined with LDA calculations or extracted by some
measurement of the Fermi surface (as for example by Angle Resolved
Photoemission Spectroscopy (ARPES)). An example of such conventional
Fermi-liquid behavior in gold has been indeed reported recently in
Ref.  \cite{Ortolani:2005:PRL}.

To make contact with the notation that we will use below while discussing
the experiments, we observe that for the Drude model \pref{drude} one
usually expresses the spectral weight in terms of the energy $\O_P^2=4\pi
ne^2/m$, which corresponds to the relations:
\be
\lb{exp}
\s_{Drude}=\frac{\O_P^2/4\pi}{\G-i\o}, \quad
W(\o_M)=\frac{\O_P^2}{4}.
\ee
Observe that in the Drude model the scale $\O_p$ is also related to
the zero of the real part of the dielectric function $\e$, which is
defined as:
\be \lb{eps} \e(\o)=\e_1+i\e_2=\e_{\infty}+\frac{i4\pi\s(\o)}{\o},
\ee
where $\e_{\infty}$ represents the screening by interband
transitions. Using Eq.~\pref{exp} we see that:
\be \lb{epsdrude}
\e_{1}^{Drude}(\o)=\e_{\infty}-\frac{\O_P^2}{\G^2+\o^2}. \ee
As a consequence, we see that when $\Gamma\ll \O_P$ the real part of the
dielectric function vanishes at a frequency $\o_0$ given by:
\be
\lb{defo0}
\e_1^{Drude}(\o_0)=0 \ra \o_0^2=\frac{\O_P^2}{\e_{\infty}}
  \left(1-\frac{\G^2}{\O_P^2}\right),
\ee
so that within the Drude model $\o_0\approx \O_P/\sqrt{\e_\infty}$
can also be related to the restricted spectral weight. Moreover,
since in conventional metals $\Gamma \ll \O_P$, in these materials the
term ``plasma edge'' refers both to the edge of intra-band optical
absorption and to the spectral weight itself of the intra-band
conductivity. However, the temperature dependence of $\o_0$ is
directly related to the temperature dependence of $\O_P$ (or $W$)
only if $\G$ is also temperature independent. Otherwise, a narrowing
of the Drude peak with decreasing temperature can lead to a blue
shift of $\o_0$ which is not necessarily related to a change of the
spectral weight $W$ \cite{Boris:2004:Science}.

\begin{figure}[htb]
\centering{
\includegraphics[width=7.cm,angle=-90]{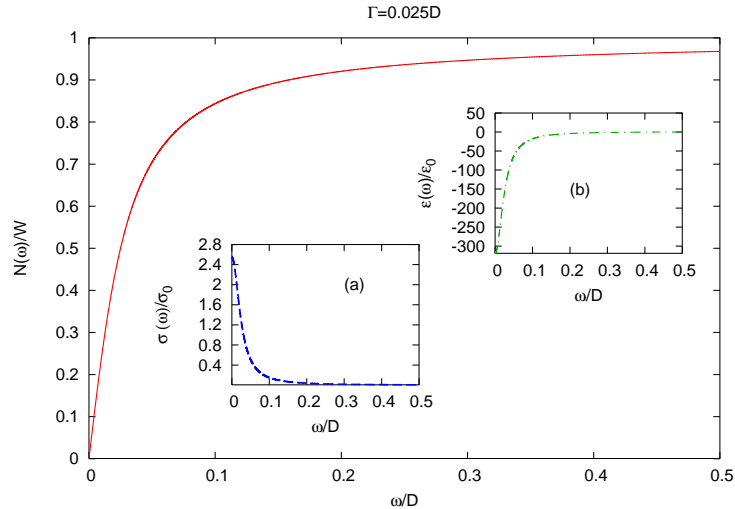}
}
\caption{Integral of the optical conductivity of the Drude model
  $N(\o)=2\int_0^\o {\rm Re}\s_{Drude}(\o')d\o'$, normalized to the total
  spectral weight $W$, as a function of $\o/D$, where $D$ is the
  bandwidth. Inset (a); plot of the optical conductivity for
  $W=0.2D$. Inset (b): real part of the dielectric function. Observe that
  here $\O_P=2W=0.4D$, where one can find the zero of the dielectric
  function. }
\label{fig-drude}
\end{figure}

\item{\bf Superconducting transition}

When a system displays a phase transition to a new ordered state, with CDW,
SDW or SC formation, the occupation number $n_{\bk\s}$ changes temperature
dependence with respect to the simple Fermi function characteristic of the
normal state of Fermi-liquid systems. A typical example is provided by the
SC transition, where within the BCS approximation one can show that the
spectral weight \pref{deft} acquires the form:
\begin{equation}
\lb{T.SC} W(\Delta,T)=\frac{2\pi e^2}{V N} \sum_{\mathbf{k}}
\frac{\partial^2 \e_\bk }{\partial k_i^2} \left[1 -
\frac{\xi_\bk}{E_\bk^{SC}} \tanh \frac{E^{SC}_\bk}{2T} \right]
\end{equation}
where $\Delta_\bk =(\Delta_0/2) (\cos k_x a - \cos k_y a)$ is the $d$-wave
SC gap and $E^{SC}_\bk = \sqrt{\xi^2_\bk + \Delta^2_\bk}$ is the
quasiparticle dispersion in the SC state. One can easily see that the
spectral weight $W(\Delta,T)$ decreases as $T$ is lowered below
$T_c$. Bearing in mind Eq.~\pref{weight.band}, this decrease can be
understood as a consequence of the kinetic-energy increase in the SC state,
where the states above $\mu$ become occupied due to the particle-hole
mixing.

\end{itemize}

\subsection{The experimental data in cuprates}

From the experimental point of view, the determination of the cut-off
$\o_M$ in Eq.~\pref{weight.band} follows naturally from the observation of
a minimum of the reflectivity (or of the conductivity) of cuprates at the
so-called {\em plasma edge} $\o_P$.  This energy scale is of the order of
$\omega_P=8000-10000$ cm$^{-1}$ in all the families of cuprates, and can be
identified as the frequency
$\o_0$ at which the real part of the dielectric function crosses zero
\cite{Ortolani:2005:PRL}.  It is then usually assumed that transitions
above $\o_P$ involve interband processes, so that $\o_P$ is a good
candidate for the cut-off $\o_M$ used in Eq.~\pref{weight.band} to relate
the spectral weight $W(\o_M,T)$ to the diamagnetic tensor, according to
Eq.~\pref{common.rule}. In this range one expects that the temperature
dependence of $W(T)$ gives information about the interactions acting on the
system.

While the identification between $\o_P$ and $\o_M$ is quite natural given
the band structure of cuprates, the optical conductivity $\s(\o)$ does not
display at all the conventional Drude-like behavior represented by
Eq.~\pref{drude}. A detailed discussion of all the peculiarities of the
optical conductivity of cuprates is beyond the scope of this paper, and can
be found for example in the recent reviews
\cite{Basov2005,review-oc,Timusk1999RPP}. We only recall here that quite
often the optical conductivity of cuprates is analyzed in terms of the
so-called extended Drude model, which relies on an extension of
Eq.~\pref{exp} including a frequency-dependent optical mass enhancement
$m^*\ra m[1+\lambda(\o)]$ and optical scattering time $\tau(\o)$, according
to the formula:
\be \lb{drudegen} \s(\o)=\frac{\O_P^2/4\pi}{1/\tau(\o)-i\o
[1+\lambda(\o)]}. \ee
Although the representations of experimental data using
$\mbox{Re}\sigma(\omega)$, $\mbox{Im}\sigma(\omega)$ and $1/\tau(\omega)$
with $1+\lambda(\omega)$ are formally equivalent, it has become rather
popular to discuss the {\em pseudogap behavior}
\cite{Basov2005,review-oc,Timusk1999RPP} in cuprate superconductors using the
language of the optical scattering rate and mass enhancement.  As we
explained above, $\O_P$ is related the reduced spectral weight $W$. Indeed,
normally $\O_P$ is found from the sum rule \pref{full.rule} written in the
form $\int_{0}^{\o_M}\mbox{Re}\sigma(\omega) = \Omega_P^2/8$. As stressed
in \cite{review-oc}, the choice discussed here of the somehow arbitrary
cut-off $\o_M$ introduces an uncertainty in the overall scale factor of the
optical scattering rate and optical mass in Eq.~\pref{drudegen}.

As one can see in Table\ I, $\O_P$ has values around 2 eV in
Bi$_2$Sr$_2$CaCu$_2$O$_8$ (Bi-2212), Bi$_2$Sr$_2$Ca$_2$Cu$_3$O$_{10}$
(Bi-2223) and YBa$_2$Cu$_3$O$_{6+x}$ (YBCO) compounds, and somewhat smaller
values in La$_{2-x}$Sr$_x$CuO$_4$ (LSCO) samples. The analysis of the
optical data is usually performed by using a cut-off $\o_M$ temperature and
doping independent, even though some variation in $\o_0$ has been reported
for example in BSCCO compounds \cite{Molegraaf:2002:Science}. However, as
stressed in Ref. \cite{Boris:2004:Science}, the temperature variations of
$\o_0$ cannot be directly related to temperature variations of the spectral
weight, because the temperature variations of $\o_0$ are also affected by
the thermal correction induced by the scattering time, see
Eq.~\pref{defo0}. In Table I we summarize some values of the
low-temperature spectral weight taken from the experiments of
Refs.~\cite{Molegraaf:2002:Science,Santander:2002:PRL, Santander:2004,
Homes:2004:PRB,Boris:2004:Science,
Ortolani:2005:PRL,Deutscher:2005,Carbone:2005}. As we explained before, the
strength of the restricted partial weight $W$ can be expressed both as
$W=\O_P^2/4$ or in terms of the kinetic energy $K$ per $Cu$ in-plane atom
as in Eq.~\pref{weight.band}.  To fix the energetic units, it can be useful
to take into account the following relations. In most of the experiments
the spectral weight $W_{exp}$ is defined as half of the value
\pref{weight.band}, because one integrates only the data at positive
frequencies. Then from Eq.~\pref{weight.band} in $n_d=2$ dimensions and
\pref{exp} we have two relations:
\be
\lb{weight.exp}
\O_P^2=4W=8W_{exp}=E_cK,
\ee
where the energy scale $E_c$ in appropriate units is:
\be
\lb{defec}
E_c=\frac{2\pi e^2 a^2}{V}= \frac{2\pi e^2}{d}=\frac{2\pi}{d}14.38 \,eV \,\AA,
\ee
and $d$ is the distance between CuO$_2$ planes. To compare the data between
single-layer and multi-layer compounds we used for $d$ the average distance
between two neighboring layers, which corresponds approximately to 13.2 \AA,
6 \AA, 7.5 \AA \ and 6.2 \AA \ in  LSCO, YBCO, Bi-2212 and Bi-2223 compounds
respectively.  Since the spectral weight $W_{exp}$ is measured usually in
$\O^{-1}$cm$^{-2}$ one should also use the equivalence 1cm$^{-1}$=0.21
$\O^{-1}$cm$^{-1}$ \cite{Homes:2004:PRB}. Then one can obtain both the
spectral weight in cm$^{-2}$ (or equivalently in eV$^2$), and the kinetic
energy $K$, given by $8W_{exp}/E_c=\O_P^2/E_c$. Finally, we note that the
background dielectric constant $\e_\infty$ has typical values around $5$
in cuprates, explaining the difference between $\o_P$ and $\O_P$ in Table\
I.

\vspace{0.5cm}
\begin{center}
\begin{tabular}{|c|c|c|c|c|c|} \hline
Ref. &Compound & $\o_M\sim \o_P$ (eV) & $\O_P(T=0)$ (eV) & $E_c$ (eV) &
$K(T=0)$ (meV)\\
 \hline
\cite{Molegraaf:2002:Science}
& Bi-2212, $T_c$=66 K, UD &  & 1.90 & 12.04& 300\\
& Bi-2212,  $T_c$=88 K, OVD & 1.2 & 2.04 &   &  346\\\hline
\cite{Santander:2002:PRL,Santander:2004,Deutscher:2005}
& Bi-2212, $T_c$=70 K, UD & 1 & 2.02& 12.04& 339\\
& Bi-2212, $T_c$=80 K, OPD &  & 1.82& & 275\\
& Bi-2212, $T_c$=63 K, OVD &  & 2.12& & 373\\ \hline
\cite{Carbone:2005}
& Bi-2223, $T_c$=110 K, OPD & 1& 2.0 & 14.61 & 276 \\\hline
\cite{Homes:2004:PRB}
& YBCO, $T_c$=57 K, UD & 1 & 1.64 & 15.05 & 180\\
& YBCO, $T_c$=91 K, OPD & 1 & 2.07 & & 285 \\ \hline
\cite{Boris:2004:Science}
& YBCO, $T_c$=92.7 K, OPD & 0.5$^*$ & 2.04 &15.05 & 277 \\ \hline
\cite{Ortolani:2005:PRL}
&LSCO, x=012, UD & 0.8 &1.41 & 6.84 & 290 \\
& LSCO, x=0.26, OVD&  &1.78 & & 463 \\ \hline

\end{tabular}\\
\hfill \\
\end{center}
Table I: Low-temperature spectral weight in several families of cuprates
and several doping (UD=underdoped, OPD=optimally doped, OVD=ovedoped). The
data at $T=0$ are either extrapolated from the data at $T>T_c$, as in
Ref. \cite{Molegraaf:2002:Science,
Santander:2002:PRL,Santander:2004,Ortolani:2005:PRL,Deutscher:2005,Carbone:2005},
or are taken at $T\approx T_c$, as in
\cite{Homes:2004:PRB,Boris:2004:Science}.  Observe that the absolute value
of the spectral weight determined in
Ref. \cite{Santander:2002:PRL,Santander:2004} is affected by an error of
15-20 $\%$, which justifies the fact that $\O_P$ it is lower in the OPD
sample than in the UD one. $^*$ (The data for YBCO extracted from
Ref. \cite{Boris:2004:Science} refer to the integral of the spectral weight
up to 0.5 eV. However, the plasma edge reported in this article is still of
order of 1 eV.)

\vspace{0.5cm}

In Table I we reported the values of the spectral weight measured at low
temperature as a function of doping. The first observation is that the
partial spectral weight increases when the doping increases. This behavior
is in agreement with the general observation that cuprate superconductors
are doped Mott insulators \cite{Toschi:2005}. Indeed, these systems behave
as if the effective number of charge carriers is proportional to the doped
holes instead of the total number of electrons, so that the spectral weight
increases when the system is doped with respect to half filling. It is
worth noting that this same behavior was observed long ago for the doping
dependence of the superfluid density \cite{Tallon:2003}, which is also
proportional to the partial spectral weight, as we shall discuss below.

While the doping dependence of the $T=0$ spectral weight is in some sense
expected for a doped Mott insulator, the recent experimental data focused
on some new interesting aspects related to the {\em temperature} dependence
of $W(T)$, both in the normal and in the superconducting state. As far as
the range of temperature $T>T_c$ is concerned, three main observations have
been made: (i) in all the compounds analyzed in Ref.
\cite{Molegraaf:2002:Science,Santander:2002:PRL, Santander:2004,
Homes:2004:PRB,Boris:2004:Science,Ortolani:2005:PRL} $W(\o_P,T)$ shows a
$T^2$ decreasing as the temperature increases, but with a coefficient
$B(\o_P)$ much larger than the one derived in Eq.~\pref{conventional} for a
standard Fermi liquid; (ii) when the cut-off frequency $\o_M$ is lower than
the plasma edge $\o_P$ the coefficient $B(\o_M)$ becomes even larger; (iii)
in underdoped compounds the opening of the pseudogap does not affect the
restricted partial weight, in particular it does not lead to any
suppression of the spectral weight, as one could expect as a consequence of
a (pseudo)gap opening.

This last observation is intimately related to the typical behavior of the
spectral weight below the SC transition, as described by the BCS relation
\pref{T.SC}. Indeed, if one assumes that the pseudogap opening is due to
some form a preformed Cooper pairing, then one could also expect to see
some decrease of the spectral weight, due to the same mechanism which leads
to the decrease of $W$ across the BCS transition. The intriguing question
arises how to reconcile the opening of the the pseudogap with an almost
conventional temperature dependence of the spectral weight, where, however,
the effect of the correlations appears in a renormalization of the
coefficient $B$ of Eq.~\pref{conventional}. To be more specific, one can
extract from the experiments the coefficient $B_{exp}$ as:
\be
\lb{normal}
W_{exp}(\O,T)=W_{exp}(\O,0)-B_{exp}(\O)T^2.
\ee
According to the relations \pref{conventional} and \pref{weight.exp}
for $\O=\o_M=\o_P$ one can relate $B_{exp}$ to the value
$B=\pi^2/12D$ of Eq.~\pref{conventional} obtained within a
tight-binding model:
\be
\lb{bexp}
B_{exp}=\frac{\pi e^2}{d}\frac{\pi^2}{24D}=\frac{E_cB}{4}.
\ee
We can then extract from $B_{exp}$ an effective hopping parameter
$t_T=\pi^2/48B$ corresponding to the thermal variation of the
spectral weight. As one can see in Table\ II, the $t_T$ values
extracted in this way are significantly smaller than the ones
obtained by ARPES measurements of the Fermi surface of cuprates,
which are usually of the order of $t\sim 300$ meV, leading to a
temperature variation $B\sim 1/t_T$ larger than expected in a
standard Fermi liquid. Even in samples were the quadratic
temperature dependence has not been clearly observed, the large
temperature variation is made apparent by analyzing the relative
spectral weight variations $\D W/W=[W(0)-W(300K)]/W(0)$, reported in
the last column of Table\ II. Observe indeed that in the usual
(non-interacting) tight-binding model $\D W/W$ never exceeds the
value of about $0.5 \%$ \cite{Benfatto:2005:PRB}. It is worth noting
that in the first works of Molegraaf et al.
\cite{Molegraaf:2002:Science} and Santander-Syro et al.
\cite{Santander:2002:PRL} on the spectral-weight behavior on
cuprates the attention was focused mainly on the absence of
spectral-weight decrease below the pseudogap temperature, but not on
the anomalously large values of $B$ measured in these BSCCO samples.
This issue was addressed for the first time from the theoretical
point of view in Ref. \cite{Benfatto:2004:EPJ}. More recently
Ortolani et al. \cite{Ortolani:2005:PRL} clearly stated this problem
in their analysis of the optical data of LSCO compounds.  As we
shall discuss in the following Sections, motivated by the
observation of the anomalous large value of $B$ we investigated
indeed the possibility to reconcile the large slope of $W(T)$ with a
DDW model for the pseudogap phase of cuprates.

\vspace{0.5cm}
\begin{center}
\begin{tabular}{|c|c|c|c|c|} \hline
Ref. &Compound & $B$ (eV$^{-1}$) & $t_T$ (meV) & $\D W/W(0)$ \\
 \hline
\cite{Molegraaf:2002:Science}
& Bi-2212, $T_c$=66 K, UD &9.3  & 21 & 0.021\\
& Bi-2212,  $T_c$=88 K, OVD & 11.3 & 18 & 0.022\\\hline
\cite{Santander:2002:PRL,Santander:2004,Deutscher:2005}
& Bi-2212, $T_c$=70 K, UD & 10.8 & 18& 0.042\\
& Bi-2212, $T_c$=80 K, OPD & 10.8 & 18&  0.053\\
& Bi-2212, $T_c$=63 K, OVD & 16.4 & 12& 0.053\\ \hline
\cite{Carbone:2005}
& Bi-2223, $T_c$=110 K, OPD & 4.2& 47 & 0.018 \\\hline
\cite{Ortolani:2005:PRL} &LSCO, x=012 & 8.3 & 24 & 0.042 \\
& LSCO, x=0.26 & 10.4 & 19 & 0.026 \\ \hline

\end{tabular}\\
\hfill \\
\end{center}
Table II: Slope $B$ of the $T^2$ temperature variation of the
spectral weight in several families of cuprates. We report here the
value obtained integrating the spectral weight up to the cut-off
$\o_M=\o_P$ as reported in Table\ I for the same compounds. The
'thermal' estimate of the hopping $t_T$ is evaluated using a flat band
dispersion, see discussion below Eq.~\pref{bexp}. Last column:
relative variation of the spectral weight between $T=0$ and $T=300$
K, $\D W/W=[W(0)-W(300K)]/W(0)$. \vspace{0.5cm}

Finally, we describe the evolution of spectral weight in the
superconducting state. When the SC order is established, part of the
low-frequency spectral weight moves into the SC response,
represented by a singular contribution to the optical conductivity:
\be
\lb{sigma.T<Tc}
\s^{SC}_{singular}(\o)=\frac{e^2n_s}{m}\left[ \pi\d(\o)+\frac{\rm
    i}{\o}\right],
\ee
where $n_s$ is the superfluid density, related to the penetration depth
$\lambda$ by the usual relation $1/\lambda^2=4\pi e^2 n_s/m$.  As a
consequence, below $T_c$ the real part of the conductivity reads:
\be \lb{distinction} {\rm Re}\s(\o) =\frac{\pi
e^2n_s}{m}\d(\o)+\sigma_{reg}, \ee
where $\sigma_{reg}$ is the regular part of the conductivity below
$T_c$. As far as the sum rule \pref{common.rule} is concerned, one
sees that the condensate contribution is always included in $W$, and
the effect of the SC transition is already taken into account by the
general formula \pref{weight.band} through the change in the
occupation number. For example, in the BCS theory the spectral
weight is expected to decrease below $T_c$, see Eq.\ \pref{T.SC}.
However, from the experimental point of view the definition
\pref{distinction} is useful to correctly determine the spectral
weight $W$ by taking into account also the condensate contribution.
Thus, one usually integrates the spectral weight of the regular part
up to $\o_{sc}$, $W_{reg}(\o_{sc})=\int_{0^+}^{\o_{sc}} {\rm
Re}\s_{reg}(\O) d\o$, obtained from the finite-frequency measurement
of the conductivity, and estimate the superfluid weight as $W_s=\pi
e^2 n_s/m$, usually using the slope of the inverse imaginary part of
$\s^{SC}_{singular}$, see Eq.~\pref{sigma.T<Tc}.  Then the value
$W=W_s+W_{reg}(\o_{sc})$ can be compared with the theoretical
prediction, which usually concerns the spectral weight integrated up
to the cut-off $\o_{sc}=\o_M$. In particular, the BCS model
\pref{T.SC} predicts a decrease of $W$ as $T$ decrease in the SC
state, to be contrasted to the increase expected in the normal state
according to Eq.~\pref{conventional}. Even though all the procedure
described above can be affected by a significant error, due mainly
to the exact determination of the superfluid weight, nonetheless it
was reported by two different group that in an underdoped and an
optimally-doped Bi-2212 compound \cite{Molegraaf:2002:Science} and
in an underdoped Bi-2212 film \cite{Santander:2002:PRL} there is an
effective {\em increase} of the spectral weight in the SC state with
respect to the value in the normal state. More specifically, it was
shown that by integrating $W_{reg}$ up to the cut-off $\o_M\sim
\o_P$ $W(T)$ is still increasing as $T$ decreases, with a
temperature variation even larger than the value $B$ found in the
normal state, see Eq.~\pref{normal}. This result has been recently
confirmed in a slightly underdoped Bi-2223 sample
\cite{Carbone:2005}. Later on, Santander et al. showed that in the
optimally-doped sample the spectral weight is almost constant in the
SC state, with $W(T<T_c)\approx W(T_c)$ \cite{Santander:2004}, as it
seems to be the case also for LSCO \cite{Ortolani:2005:PRL}.
Finally, a more conventional result has been observed in an
overdoped Bi-2212 film \cite{Deutscher:2005}, showing that the
anomalous spectral-weight behavior can be eventually doping
dependent. However, these results has been questioned in Ref.
\cite{Boris:2004:Science}, where the optical data of an
optimally-doped YBCO sample and slightly-underdoped Bi-2212 sample
up to a larger cut-off of $\sim 1.5$ eV were considered. According
to the analysis of Boris et al., the spectral weight of YBCO and
Bi-2212 in this frequency range keeps {\em constant} in the normal
state and decrease when the SC state is formed. Thus, the issue of
the spectral-weight behavior below $T_c$ remains at the moment quite
controversial \cite{Kuzmenko:2005,Santander:2005}, and seems to be
related mainly to the analysis of the optical data rather than to an
effective discrepancy between them. In particular, it seems to us
that in Ref.~\cite{Boris:2004:Science} it is mainly questioned the
choice of the plasma edge $\o_P\sim 1$ eV as the proper cut-off
$\o_M$ for the analysis of optical data. However, this is the most
plausible choice from the theoretical point of view, because
extending the spectral-weight integration to interband processes one
naturally recovers the full f-sum rule \pref{full.rule}, which is
just a constant and does not provide any significant information on
the interactions acting on the system.

\section{Optical conductivity sum rule for a reduced $d$-density wave model}
\lb{sec:reduced_model}

While it is clear that any successful theory of HTSC should be able
to explain the above-mentioned experimental facts, despite the 20
year jubilee since the discovery of HTSC, there is no consensus yet
neither on the general theory of HTSC, nor on the particular
explanation of these facts.

Although our goal is to present here the latest theoretical results
for the optical-conductivity sum rule within the {\em $d$-density
wave\/} paradigm, we would like to mention some other works related
to this issue in the recent literature. A first class of papers is
related to the spectral-weight below the superconducting critical
temperature. For example, the possibility of a spectral-weight
change below $T_c$ in terms of the lowering of the in-plane kinetic
energy has been analyzed in
Refs.~\cite{Hirsch:2000:PRB,Norman:2002:PRB,Kopec2003PRB,Eckl:2003:PRB}.
In Ref.~\cite{Hirsch:2000:PRB} a model with ``occupation modulated''
hopping terms was considered. The reduction of the kinetic energy at
$T_c$ was attributed in Refs.~\cite{Kopec2003PRB,Eckl:2003:PRB}) to
the transition from a phase-incoherent Cooper pair motion in the
pseudogap regime above $T_c$ (see, e.g. review
\cite{Loktev:2001:PRP}) to a phase coherent motion at $T_c$, while
in Ref.~\cite{Norman:2002:PRB} a model with a frequency dependent
scattering rate was used. More recently, the optical-conductivity
sum rule was analyzed for a model with electrons in the finite band
coupled to a single Einstein oscillator \cite{Knigavko:2004:PRB} and
in the nearly antiferromagnetic Fermi liquid model
\cite{Schachinger:2005:PRB}.

As far as the anomalous behavior of the spectral weight in the normal-state
is concerned, a possible interpretation was recently given by Toschi {\em
et al.} \cite{Toschi:2005}. In this work the authors evaluate the
spectral-weight behavior within the dynamical mean-field analysis of the
repulsive Hubbard model, and they show that the large slope $B$ (see
Eqs.~\pref{conventional} and \pref{normal}) is a consequence of the
bare-bandwidth renormalization $B\sim 1/Zt$ where $Z\rightarrow 0$ as the
Mott-insulator is approached. At the same time the $T=0$ value of $W$ is
much less affected by the $Z$ renormalization, due to the contribution of
the so-called mid-infrared processes, which occur below $\o_P$ and partly
compensate for the reduction of the quasi-particle weight
\cite{Toschi:2005}.

Finally, the behavior of the spectral weight in the DDW state was addressed
specifically in a series of recent papers, Ref.
\cite{Benfatto:2004:EPJ,Benfatto:2005:PRB,Valenzuela:2005:PRB,Aristov:2005,Gerami:2005}.
While the issue of the vertex corrections, discussed below, is the most
important ingredient of the Refs.
\cite{Benfatto:2004:EPJ,Benfatto:2005:PRB,Aristov:2005}, in
Ref.~\cite{Valenzuela:2005:PRB} they were ignored and the main attention
was paid to the analysis of the frequency dependence of $\sigma(\omega)$
both in the presence and in the absence of the next-nearest-neighbor
hopping. As we will see in Sec.~\ref{sec:full-model}, the results of
Ref.~\cite{Valenzuela:2005:PRB} are particularly useful for the discussion
of interband optical transitions when the vertex correction becomes
unimportant. At the same time, Ref.~\cite{Gerami:2005} investigates the
consequences of a momentum-dependent scattering time on the optical
spectra, neglecting again the contribution of vertex corrections.

\subsection{Unconventional charge density waves}

Since 2000 the most commonly used name for the state that we
consider in the present article is {\em $d$-density wave (DDW)
state\/} \cite{Chakravarty:2001:PRB}. It is considered as one among
competing theoretical approaches that are attempting to explain the
pseudogap behavior in cuprates \cite{Timusk1999RPP}. However, we
{\em do not} intend to analyze here to which extent the DDW model
can explain the vast body of the experimental data accumulated
during the last 20 years.  Instead of this we will concentrate on
rather important theoretical subtleties of the DDW model that are
often overlooked and their implications for the optical
conductivity. Yet before writing down the model Hamiltonian we find
useful to remind the reader of the rather long history of the model.

The DDW or, as it was called originally, the {\em orbital
antiferromagnetic state\/} was introduced in solid state physics in
Ref.~\cite{Halperin:1968:SSP}. The discovery of HTSC and searches
for a model that would describe the CuO$_2$ planes without phonons
lead to the investigation of the Heisenberg-Hubbard model
\cite{Affleck:1988:PRB,Marston:1989:PRB,Nersesyan:1989:JLTP,Schulz:1989:PRB}
and to the rediscovery of this state which at this time was called a
{\em staggered flux phase\/}. The phase {\em with circulating
orbital currents\/} turned out to be an attractive idea and there
are hundreds of papers (see, for instance, very few of them in Refs.
\cite{Varma:1997:PRB,Eremin:1998:JETPL,Cappelluti:1999:PRB,Benfatto:2000:EPJ,Maki})
that exploit and develop it. These studies showed that the DDW state
can be obtained when the interaction term in (\ref{Hamiltonian}),
for example, is  Heisenberg
\begin{equation}
\lb{Heisenberg} H_{int} = \frac{J}{2} \sum_{\substack{{<ij>}\\
\alpha \beta \gamma \delta }} c^\dagger_{i\alpha}
\vec{\sigma}_{\alpha \beta} c_{i\alpha} c^\dagger_{j\gamma}
\vec{\sigma}_{\gamma \delta} c_{j \delta},
\end{equation}
or Hubbard-like
\begin{equation}
\lb{Hubbard} H_{int} = \sum_{ij,\s\s'}
c_{i\sigma}^{\dagger}c_{i\sigma} V(\mathbf{r}_i -
\mathbf{r}_j)c_{j\sigma'}^{\dagger}c_{j\sigma'}.
\end{equation}
The sum in Eq.~(\ref{Heisenberg}) runs over nearest neighbors sites
$i$ and $j$, $\vec{\sigma}$ is the vector of Pauli matrices and
$V(\mathbf{r}_i - \mathbf{r}_j)$ in Eq.~(\ref{Hubbard}) is a generic
density-density interaction. For example, the repulsive Hubbard
model corresponds to the case $V(\br_i-\br_j)=U$ for $i=j$.

In the mean-field treatment the expectation value of any two operators,
i.e. the order parameter, may be nonzero. In particular, if the expectation
value of two operators with the same spin projection is nonzero we deal
with a charge-density wave (CDW), which can have an internal symmetry if
the two operators in the order parameter belong to different sites. CDW
with an internal symmetry are called unconventional charge-density waves to
distinguish them from conventional ones, where the electron and hole sit on
the same site. An other possibility is that the order parameter describes
microscopic currents running on few sites of the lattice, as it is the case
for the flux phase
\cite{Affleck:1988:PRB,Marston:1989:PRB,Nersesyan:1989:JLTP,Schulz:1989:PRB}
or for the current patterns proposed by Varma \cite{Varma:1997:PRB}.  It is
worth noting that strictly speaking the flux phase does not present a
charge modulation, even though a gap is present in the excitation
spectrum. Indeed, the same phenomenological spectrum of the DDW state has
been proposed as emerging in cuprates due to the tendency of the system to
form charge order. This attitude was for example considered in
Ref.~\cite{Benfatto:2000:EPJ}, where the DDW state emerges as the
consequence of the proximity of the system to a quantum critical point
\cite{Varma:1997:PRB,Castellani}. More generally, unconventional charge
density waves have been observed in many other systems, as for example
organic conductors \cite{Maki} or dichalcogenides
\cite{Vescoli:1998:PRL,Neto:2001:PRL}, but we will restrict our discussion
here to cuprate superconductors. In the following discussion it will be
clear that the crucial point in the evaluation of the optical conductivity
for a state with DDW or, more general, with a CDW with internal symmetry,
is the non-trivial dependence of the electronic self-energy on the
momentum.  For this reason the results reported below are generic to many CDW
systems, even though we will consider specifically the case of DDW order in
cuprate superconductors.

\subsection{Reduced $d$-density wave model}

In the DDW case, the general interaction term \pref{Hubbard}  is
replaced by an approximate reduced interaction Hamiltonian:
\be
\lb{int} H_I=-\frac{V_0}{2N}\sum_{{k,k'\atop{\s,\s'}}} w_d(\bk)
w_d(\bk') c_{\bk\sigma}^\dagger c_{\bk+\bQ\sigma}
c_{\bk'+\bQ\sigma'}^\dagger c_{\bk'\sigma'},
\ee
where $w_d(\bk)=(\cos k_xa-\cos k_ya)/2$ corresponds to $d$-wave
symmetry with respect to the discrete rotation group of the square
lattice. By defining  $iD_0= -(V_0/N)
\sum_{\bk\s}w_d(\bk)<c_{\bk+\bQ\sigma}^\dagger c_{\bk\sigma}>$ and
adding the non-interacting Hamiltonian $H_0$ from \pref{Hamiltonian}
we arrive at the {\em mean-field} DDW Hamiltonian
\begin{equation}
\lb{DDW-Hamiltonian} H =  \sum_{\mathbf{k},\sigma} [(\varepsilon_\bk
-\mu) c_{\bk\sigma}^\dagger c_{\bk\sigma} + i D_\bk
c_{\bk\sigma}^\dagger c_{\bk+\bQ\sigma}],
\end{equation}
where $D_\bk = D_0 w_d(\bk)$ is the gap, known as the DDW gap
\cite{Chakravarty:2001:PRB}, arising from the formation of the state
with circulating currents below a characteristic temperature
$T_{DDW}$. In writing Eq.~(\ref{DDW-Hamiltonian}) we considered the
case of $t'=0$ in the bare band dispersion, so that the nesting
condition $\varepsilon_{\mathbf{k} + \mathbf{Q}} = -
\varepsilon_{\mathbf{k}}$ is satisfied, where $\mathbf{Q} = (\pi/a,
\pi/a)$ is the characteristic wave-vector at which a particle-hole
coupling in the DDW state is considered. However, the following
results are easily extended to the case $t'\neq 0$ (see, for
example,
Refs.~\cite{Aristov:2004:PRB,Benfatto:2005:PRB,Valenzuela:2005:PRB}).

\begin{figure}[htb]
\centering{ \includegraphics[width=7.cm, angle=-90]{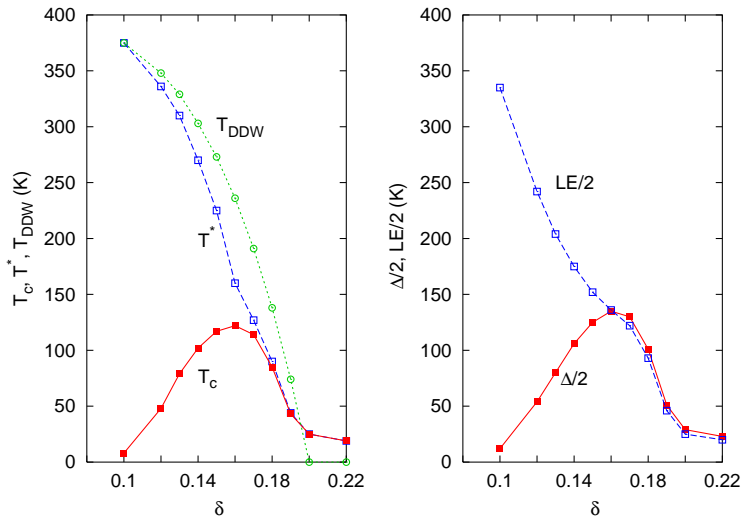}
} \caption{Phase diagram of cuprates [taken from
Ref.~\cite{Benfatto:2000:EPJ}] derived adding the BCS pairing term
\pref{H.BCS} to the Hamiltonian \pref{DDW-Hamiltonian}. $T_{DDW}$ is the
temperatures where $D_0$ forms, $T^*$ is the temperature where a
leading-edge (LE) shift appears in the quasiparticle spectrum at the
$M$ points, and $T_c$ is the SC temperature, at which the SC gap
$\D$ forms.} \label{fig:diagram}
\end{figure}

The notation is then simplified by halving the Brillouin zone and
introducing two-component electron operators (the DDW equivalent of
Nambu spinors \cite{Nambu:1960:PR} widely used in the theory of
superconductivity \cite{Schrieffer})
\be \lb{spinors}  \chi_{\bk \s}^{\dagger} = \left(
\begin{array}{cc} c_{\bk \s}^{\dagger} \quad c_{\bk +
\mathbf{Q},\s}^{\dagger}
\end{array} \right),
\ee
where $c_{\bk \s}^{\dagger}$ and $c_{\bk \s}$ are the Fourier
transforms of $c_{i\s}^{\dagger}$ and $c_{i\s}$. The Hamiltonian
(\ref{DDW-Hamiltonian}) written in terms of $\chi$ becomes
\begin{equation}
\lb{DDW-Hamiltonian-matrix} H = \sum_{\mathbf{k},\sigma}^{RBZ}
\chi_{\bk \s}^{\dagger} \left[ \varepsilon_\mathbf{k} \sigma_3
-D_\bk \sigma_2- \mu \right] \chi_{\bk \s},
\end{equation}
where the sum is taken over the reduced (magnetic) Brillouin zone
(RBZ). Thus one can easily see that when $D_0\neq 0$ two quasiparticles
excitation branches are formed with dispersion $\xi_{\pm,\bk}=-\mu\pm
E_\bk$, where $E_\bk=\sqrt{\e_\bk^2+D_\bk^2}$. As a consequence, at finite
doping with respect to half-filling the Fermi surface consists of small
pockets around the $(\pm\pi/2,\pm\pi/2)$ points and a finite leading-edge
shift $LE\approx D_0-\mu$ at the $M$ points, which mimic in some sense the
arc of Fermi surface observed by ARPES measurements in the underdoped phase
of cuprates \cite{Benfatto:2000:EPJ,Valenzuela:2005:PRB}.  For this reason,
the Hamiltonian \pref{DDW-Hamiltonian} can be considered as an
``effective'' model for the normal (i.e. non SC) pseudogap phase of
cuprates, where SC forms within a standard BCS mechanism, since all the
anomalies related to the presence of a pseudogap are already included in
the Hamiltonian \pref{DDW-Hamiltonian}. This point of view was adopted for
example in Ref. \cite{Benfatto:2000:EPJ}, where the phase diagram of
Fig.~\ref{fig:diagram} was derived, in excellent agreement with the
experimental data on BSCCO.

%%%%%%%%%%%%%%%%%%%%%%%%%%%%%%%%%%%%%%%%%%%%%%%%%%%%%%%%%

%%%%%%%%%%%%%%%%%%%%%%%%%%%%%%%%%%%%%%%%%%%%%%%%%%%%%%%%%

If one consider the Hamiltonian \pref{DDW-Hamiltonian} as an effective
model, it is straightforward to calculate the current-current correlation
function and the electrical conductivity for it, because it describes
non-interacting quasiparticles.  Nevertheless, one should be careful and
take into account that the current operator and the diamagnetic tensor for
the Hamiltonian \pref{DDW-Hamiltonian} are different from those
given by Eqs.~(\ref{defj}) and (\ref{deft}) for the Hamiltonian
\pref{Hamiltonian}.

This can be understood by deriving the particle current operator
compatible with the charge conservation law  and with the equations
of motion for the operators $c$ and $c^\dagger$,
\cite{Yang:2002:PRB,Kee:2002:PRB,Sharapov:2003:PRB,Aristov:2004:PRB}
\begin{equation}
\lb{electric.current.DDW} \mathbf{j}
(\mathbf{q},t)=\frac{1}{N}\sum_{\mathbf{k}, \sigma} \left[ v_\bk^F
c^{\dagger}_{\mathbf{k}- \mathbf{q}/2\s}
c_{\mathbf{k}+\mathbf{q}/2\s} -iv_\bk^D c^{\dagger}_{\mathbf{k}-
\mathbf{q}/2\s} c_{\mathbf{k} + \mathbf{Q}+ \mathbf{q}/2\s} \right],
\end{equation}
where $v^D_{\bk_i}=-\pd D_\bk/\pd\bk_i$ is the DDW gap velocity. The first
term of the previous expression relates as usual the particle
current to the band velocity $v_\bk^F$, see Eq.~\pref{vbare}. The
second term, which only appears for non-vanishing $D_0$, takes into
account the contribution of the orbital currents to the electrical
conductivity, and it emerges when the DDW order develops. Observe
that in the reduced model \pref{DDW-Hamiltonian} the term
proportional to $D_0$ appears as an additional, temperature
dependent band, which couples $\bk$ and $\bk+\bQ$ electrons, and as
a consequence a corresponding term appears in the definition of the
current. By rewriting the electric current operator
\pref{electric.current.DDW} using the spinors \pref{spinors}, one
has
\begin{equation}
\lb{electric.current.DDW-spinor} j_i (\mathbf{0},t)
 = \frac{1}{N}\sum_{\bk \sigma}^{\mathrm{RBZ}}
\chi_{\mathbf{k} \s}^{\dagger}  V_i(\mathbf{k})
 \chi_{\mathbf{k} \s},
\end{equation}
where $V_i(\bk)$ is a generalized velocity which extends the definition
\pref{vbare} to the case of Hamiltonian \pref{DDW-Hamiltonian}
\be
\lb{Vi}
V_i({\bk }) \equiv v^F_{{\bf k}_i}\s_3+v^D_{{\bf
k}_i}\s_2.
\ee
Note that this result appears to be rather trivial, if one considers
the original paper \cite{Affleck:1988:PRB}, where the staggered flux
phase was introduced, and the sum of hopping and staggered flux terms
was considered as the ``effective'' hopping term.

The modification induced by the DDW order parameter on the current
operator demands a consistent change in the diamagnetic tensor with
respect to the definition \pref{deft}. As we discussed in
Sec.~\ref{sec:general-sigma-sum}, after the Peierls substitution
both the current operator and the diamagnetic tensor can be derived
from $H(A)$, according to Eq.~\pref{j-tau.def}. As a consequence, by
performing the Peierls substitution in the reduced model
\pref{DDW-Hamiltonian} not only the current operator but also the
diamagnetic tensor $\langle \tau_{ii} \rangle$ is modified,
containing an extra term for $D_0\neq 0$ \cite{Benfatto:2004:EPJ},
\be
\lb{new.tensor}
\langle \t_{ii}\rangle=-\frac{a^2}{2N}\sum_{\bk\s}
\left[\e_\bk \langle c_{\bk\s}^\dagger c_{\bk\s} \rangle+ i D_\bk
\langle c_{\bk\s}^\dagger c_{\bk+\bQ\s}\rangle \right].
\ee
When the operator averages are evaluated (see
Refs.~\cite{Benfatto:2004:EPJ,Benfatto:2005:PRB} for the details),
one finds that the sum rule for the reduced model is:
\begin{equation}
\lb{sum.T>Tc} \frac{W^{DDW}(D,T)}{(\pi e^2a^2/V)}=-
\frac{1}{N}\sum_{\bk}^{RBZ} E_\mathbf{k} [f (\xi_{+,\mathbf{k}}) -
f(\xi_{-,\mathbf{k}})].
\end{equation}
Eq.~(\ref{sum.T>Tc}) was derived using the fact that $\partial_{x,y}
v^F_\bk= 2ta^2 \cos k_{x,y} a$ (and $\partial_{x,y} v^D_\bk= \pm
(D_0/2) a^2 \cos k_{x,y}a$), and it reduces to
Eq.~(\ref{weight.band}) for $D_0=0$. Another way to derive the
result (\ref{sum.T>Tc}) is to evaluate the commutator of the density
$j_0(\mathbf{q},t)$ and the current $j_i(\mathbf{q},t)$, given by
Eq.~\pref{electric.current.DDW}, as explained in
Refs.~\cite{Benfatto:2004:EPJ,Benfatto:2005:PRB}.

As we mentioned above, one can consider the superconductivity formation on top
of the DDW state as arising from a conventional BCS type of pairing.
More specifically, we can add to the Hamiltonian
(\ref{DDW-Hamiltonian-matrix}) an additional d-wave mean-field
pairing term
\be
\lb{H.BCS}
H_p= \sum_{\mathbf{k}} [\Delta^{\ast}_\bk
c_{-\bk\downarrow}c_{\bk\uparrow} + h.c.], \ee
where the SC $d$-wave gap $\Delta_\bk$ is defined after
Eq.~\pref{T.SC}. In the SC phase the occupation numbers of
Eq.~\pref{new.tensor} get modified, and the spectral weight in the
DDW+SC state reads
\begin{equation}
\lb{sum.complete}
\frac{W^{DDW}(D,\Delta,T)}{\pi e^2 a^2}= \frac{1}{2V
N}\sum_{\mathrm{RBZ}} E \left[\frac{\xi_{+}}{E_+} \tanh
\frac{E_+}{2T} - \frac{\xi_{-}}{E_-} \tanh \frac{E_-}{2T} \right],
\ee
where $E_{\pm,\mathbf{k}}=\sqrt{\xi_{\pm,\bk}^2+\Delta^2_\mathbf{k}}$ is
the quasiparticle dispersion in the presence of pairing, and the explicit
dependence on $\bk$ has been omitted.

\subsection{Results for the reduced model}

The numerical calculation of the spectral weight defined by
Eqs.~\pref{sum.T>Tc} and \pref{sum.complete} was done in
Ref.~\cite{Benfatto:2004:EPJ} and the results at doping $\d=0.13$
are reported in Fig.~\ref{fig:EPJ}. As one can see, below $T_{DDW}$
the spectral weight $W^{DDW}$ defined in Eq.~\pref{sum.T>Tc} is
larger than the value obtained in the absence of the DDW order,
represented by the dotted-dashed line in Fig.\ \ref{fig:EPJ}. When
also SC is established, the spectral weight $W^{DDW}(D,\D)$ given by
Eq.\ \pref{sum.complete} is slightly smaller than in the DDW state
only, the change being proportional to the ratio $\D/D_0$ which is
quite small at this doping. In the inset of Fig.~\ref{fig:EPJ} we
plot $W(T)$ as a function of $T^2$.  One can see that below
$T_{DDW}$ the $T^2$ temperature dependence of $W(T)$ is still
recovered over a wide range of temperature, but with a slope much
larger than in the non-DDW case. All these results show some
interesting resemblance to the experimental data for cuprates
discussed in Sec.~II-C. In particular, it turns out that the large
spectral-weight variation measured by the experiments of
Ref.~\cite{Molegraaf:2002:Science,Santander:2002:PRL,Santander:2004,Ortolani:2005:PRL,Deutscher:2005,Carbone:2005}
between $T=0$ and $T=300$ K$\sim 0.1t$ can be ascribed to existence
of DDW order in the normal state. Moreover, when SC is formed on top
of this DDW state the spectral weight remains almost constant, as
observed in Ref. \cite{Santander:2004}. However, one should notice
that on the overdoped side, the temperature $T_{DDW}$ of DDW
formation decreases (see Fig.\ \ref{fig:diagram}) so that $T_{DDW}$
can be smaller than room temperature, and a change of slope of the
spectral weight at $T_{DDW}$ should be observed. Up to now, no
signature of this effect has been observed in the experiments.

%%%%%%%%%%%%%%%%%%%%%%%%%%%%%%%%%%%%%%%%%%%%%%%%

\begin{figure}
\centering{ \includegraphics[width=9.cm, angle=0]{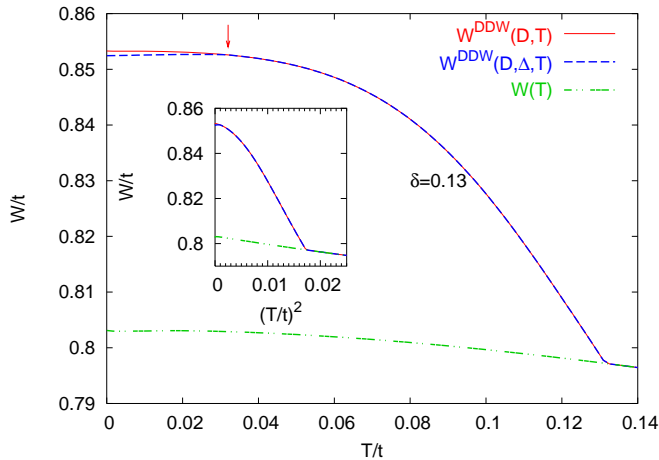} }
\caption{Spectral weight in units of $e^2\pi a^2/V$ in the normal state
Eq.~\pref{weight.normal} (dot-dashed line), in the DDW state
Eq.~\pref{sum.T>Tc} (dashed line) and in DDW+SC state
Eq.~\pref{sum.complete} (full line).  The values of parameters are the same
used in Fig.~\ref{fig:diagram} for the doping $\delta=0.13$ ($D_0(T=0)=0.92
t$, $\Delta(T=0)=0.064t$, see Ref.\
\cite{Benfatto:2000:EPJ,Benfatto:2004:EPJ} for further details). The
critical temperature is marked by the arrow. Observe that the decrease of
$W^{DDW}(D,\Delta,T)$ below $T_c$ is almost negligible. Inset: spectral
weight plotted as function of $(T/t)^2$.}
\label{fig:EPJ}
\end{figure}

%%%%%%%%%%%%%%%%%%%%%%%%%%%%%%%%%%%%%%%%%%%%%%%%%

To develop a better insight into the origin of the spectral-weight
increase in the DDW state we calculated in
Ref.~\cite{Benfatto:2005:PRB} the optical conductivity corresponding
to the sum rule \pref{sum.T>Tc}. This can be done using the general
definition \pref{conductivity.def} where, however, the correlation
function \pref{cur-cur} is modified according to the new current
definition \pref{electric.current.DDW}. Let us first introduce the
Green's function (GF) corresponding to the effective Hamiltonian
(\ref{DDW-Hamiltonian-matrix}):
\be \lb{Green-DDW} G^{-1}(\mathbf{k}, i\o_n)=  (i\omega_n + \mu)\s_0
-\e_\bk\s_3+D_\bk\s_2. \ee
Using this GF one can write down an {\em exact\/} expression for the
correlation function
\be \lb{pigi-DDW} \Pi_{ij}(\bq=\mathbf{0},i\O_m)=
-2\frac{T}{N}\sum^{RBZ}_{\bk,i\o_n} \mbox{Tr}[G(\bk,i\o_n+i\O_m)
\gamma_i(\bk)G(\bk,i\o_n)\gamma_j(\bk)], \ee
where
\be \lb{gstatic} \gamma_i(\bk)= V_i(\bk), \ee
with $V_i(\bk)$ defined in Eq.~(\ref{Vi}) is the bare vertex for our
noninteracting theory \cite{note-vertex}. One can also convince
oneself that \pref{gstatic} is indeed the appropriate vertex for the
reduced Hamiltonian \pref{DDW-Hamiltonian} by looking at the Ward
identity (see Eq.~\pref{WI-general} below)
\be \lb{Ward-DDW} \gamma_i(\bk) = -\frac{\partial
G^{-1}(\bk,i\omega_n)}{\partial \bk_i} \ee
which is just the charge-conservation law expressed in terms of the Green's
and vertex functions.  Then one can state that the term $\sim \sigma_2
v_\bk^D$ in the vertex \pref{gstatic} is related to the $\bk$-dependence of
the DDW gap in the GF \pref{Green-DDW}.  Using the definition
\pref{pigi-DDW} of the correlation function and adding an isotropic
impurity scattering we evaluated in Ref.~\cite{Benfatto:2005:PRB} the
optical conductivity in the DDW state, see the inset of
Fig.~\ref{fig:conductivity}. Here we report the $T=0$ value of the optical
conductivity in a system without DDW and with DDW for $\d=0.13$, using the
same parameter values as in Fig.~\ref{fig:EPJ}. As usual, when DDW order
forms the optical conductivity presents both a Drude peak and an absorption
at finite frequency $\sim 2\mu$ due to the optical processes between the
two quasiparticle branches $\xi_\pm$ (see also Ref.
\cite{Valenzuela:2005:PRB} for the case with $t'\neq 0$).  Although the
Drude peak is reduced by the DDW formation, these interband
finite-frequency processes lead to an overall increase of the spectral
weight with respect to the case without DDW, as one can see in the main
panel of Fig.\ \ref{fig:conductivity} where we report $N(\o)=2\int_0^\o
\s(\o')d\o'$. The crossing of $N^{DDW}$ with respect to $N$ at $\o\sim 2t$
is again an effect of the additional term proportional to $v_D$ in the
current \pref{electric.current.DDW} and in the vertex \pref{gstatic}, which
extend above the Drude peak the effect of the DDW formation. Indeed, as
discussed in detail in Ref.~\cite{Benfatto:2005:PRB}, and consistent with
the results of Valenzuela {\em et al.} in Ref.~\cite{Valenzuela:2005:PRB},
when the contribution proportional to $v_D$ is omitted one finds instead in
the DDW state a spectral weight lower than in the normal state (in analogy
with the behavior across the SC transition described by ordinary BCS
theory).

%%%%%%%%%%%%%%%%%%%%%%%%%%%%%%%%%%%%%%%%%%%%%%%%

\begin{figure}[htb]
\centering{ \includegraphics[width=6.cm, angle=-90]{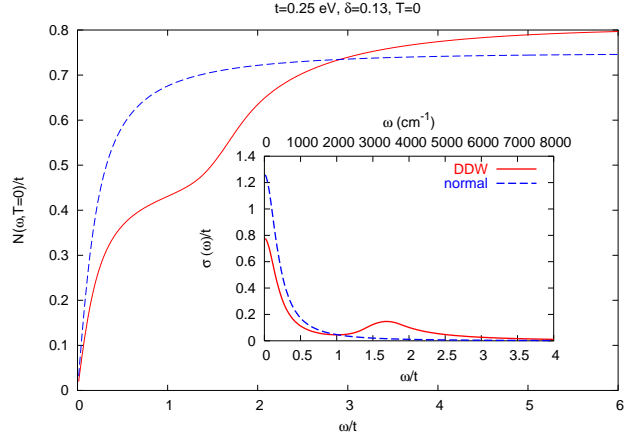}
} \caption{Integrated spectral weight in units of $e^2\pi a^2/V$ for
the optical conductivity at $T=0$ of a system with and without DDW.
The values of parameters are the same as in Fig.~\ref{fig:EPJ}. At
low cut-off energy the spectral weight in the DDW state is lower
than in the normal state, but when intraband excitations are taken
into account the spectral weight of the DDW state exceeds the value
of the normal state. Inset: optical conductivity in the two cases.}
\label{fig:conductivity}
\end{figure}

%%%%%%%%%%%%%%%%%%%%%%%%%%%%%%%%%%%%%%%%%%%%%%%%%

\section{Optical conductivity and sum rule within a full model
with DDW instability}
\lb{sec:full-model}

\subsection{Limitations of the reduced model and need of a more microscopic
approach}

There is no doubt that the results presented above for the electric current
operator \pref{electric.current.DDW}, the optical conductivity shown in
Fig.~\ref{fig:conductivity} and the sum rule \pref{sum.T>Tc} (and its
generalization \pref{sum.complete} valid in the superconducting state that
forms on top of the DDW state) are consistent with each other, as far as
they are considered within the framework of the reduced model
\pref{DDW-Hamiltonian}.

One could, however, still question these results from the more general
point of view of the validity of the Hamiltonian \pref{DDW-Hamiltonian},
viz. one may question whether the low-energy Hamiltonian
\pref{DDW-Hamiltonian} is suitable for the description of the intraband
excitations that cause the increase of $W^{DDW}(D,T)$ as $T$ decreases. The
justification of the Hamiltonian \pref{DDW-Hamiltonian} can only be made
using a more microscopical Hamiltonian \pref{Hamiltonian}. In many ways
this resembles the history of the reduced BCS Hamiltonian, since the theory
of superconductivity was completed only when a more generic Hamiltonian was
studied (see e.g.  Refs.~\cite{Schrieffer,Nambu:1960:PR}). These analogies
between DDW and BCS models were exploited in
Ref.~\cite{Benfatto:2005:PRB}. One of the main conclusions of
Ref.~\cite{Benfatto:2005:PRB} is that the relation between the optical
conductivity $\sigma(\omega)$ and the sum rule for the full model
(\ref{Hamiltonian}) can be established by looking at the problem of the
vertex corrections in the current-current correlation functions. This
vertex is determined by solving an integral equation, whose solution is a
formidable problem. Its zero-frequency solution provided by the Ward
identity was used in Ref.~\cite{Benfatto:2005:PRB}. More recently Aristov
and Zeyher ~\cite{Aristov:2005} extended this analysis to finite frequency
and investigated the behavior of sum rule in the full DDW model.

\subsection{Electrical conductivity and equation for vertex}

The current-current correlation function defined by
Eq.~\pref{cur-cur} is expressed in terms of the DDW Green's function
\pref{Green-DDW}, the bare vertex $\gamma^0_i$ and the full vertex
function $\Gamma_j$ as follows \cite{Schrieffer}:
\be \lb{pigi-full} \Pi_{ij}(\bq=\mathbf{0},i\O_m)=
-2\frac{T}{N}\sum^{RBZ}_{\bk,i\o_n} \mbox{Tr}[G(\bk,i\o_n)
\gamma^0_i(\bk)G(\bk,i\o_n+i\O_m)\Gamma_j(\bk,i\O_m)]. \ee
where $\gamma^0$ and $\Gamma$ represent the bare and full vertex function,
respectively. A careful reader may notice that the vertex $\gamma^0_i$ for
a model based on the full Hamiltonian \pref{Hamiltonian} is nothing else
then Eq.~(\ref{vbare}). Using the matrix notation introduced in
Sec.~\ref{sec:reduced_model} it reads
\begin{equation}
\lb{bar-vertex} \gamma^0_\mu(\bk)=( v_{\bk}^F\s_3, \s_0) .
\end{equation}
Comparing the bare vertex $\gamma_i^0$ with the vertex $\gamma_i$
[see Eqs.~\pref{gstatic} and \pref{Vi}] of the reduced model, it
becomes clear that the bare bubble approximation when the full
vertex $\Gamma_j(\bk,\omega)$ is replaced at will by the bare vertex
$\gamma^0_i(\bk)$ would miss completely the physics of the problem.
Thus one has to find the full vertex by solving the integral
equation for it \cite{Benfatto:2005:PRB,Aristov:2005}
\begin{equation}
\lb{vertex-eq} \Gamma_{i} (\bk, i \Omega_m) =
       \gamma_{i}^0(\bk)
         + T \sum_{x_l}\int d\bp
       V(\bp- \bk) G (\bp, i x_l)
       \Gamma_{i} (\bp, i \Omega_m)
       G(\bp, i x_l +  i \Omega_m),
\end{equation}
where  the current vertex $\Gamma_i$ is a $2\times 2$ matrix and the
integration over $\bp$ with the measure $d \bp = d p_x d p_y/(2
\pi)^2$ runs over the full (chemical) Brillouin zone, while $\bq$
lies within the reduced zone. Here $V(\bp-\bq)$ represents the
Fourier transform of the potential $V(\br_i-\br_j)$ of
Eq.~\pref{Hubbard}. In Ref.~\cite{Aristov:2005} the interaction term
of the Hamiltonian $H_{int}$ is chosen to be the Heisenberg
interaction \pref{Heisenberg} between nearest neighbors, so that in
the momentum representation $V(\bk)=J(\bk) = 2 J(\cos k_x a + \cos
k_y a)$. After analytical continuation $i \Omega_m \to \omega+i0$ is
made, one can consider the static limit $\omega=0$ of
Eq.~\pref{vertex-eq} to illustrate that if a vertex function
$\Gamma_i$ satisfies the Ward identity
\be \lb{WI-general}\Gamma_i({\bk},0) = -\frac{\partial
G^{-1}(\bk,0)}{\partial \bk_i} \ee
it is also a solution of the equation \pref{vertex-eq} for vertex.
The crucial assumption in this proof (see
Refs.~\cite{Schrieffer,Benfatto:2005:PRB}) is that the potential $V$
is nonseparable, viz. it depends on the difference $\bp - \bk$. The
restrictions on the form of the potential $V(\bp,\bk)$ necessary to
satisfy the Ward identity \pref{WI-general} and, respectively, to
obey the charge conservation law deserve special study (see e.g.
Refs.~\cite{Vollhardt:1980:PRB,Nieh:1998:PLA}). It was observed in
Ref.~\cite{Aristov:2005} that an interesting feature of
Eq.~\pref{vertex-eq} is that, for the specific potential
$J(\bp-\bk)$ considered there, only the odd kernels (like $\sin p_x
a \sin k_x a$) of its decomposition into the sum of separable
kernels can contribute in the integral over $\bp$, while these terms
were absent in the reduced model \pref{int}. Finally, we note that
the relation \pref{WI-general} was used above in
Eq.~\pref{Ward-DDW}, since for the reduced non-interacting model
\pref{DDW-Hamiltonian} it allows to obtain the exact vertex valid
for all frequencies \cite{Benfatto:2005:PRB}.

The equation for the vertex \pref{vertex-eq} with the potential
$J(\bk-\bp)$ written above is solved at finite frequency in
Ref.~\cite{Aristov:2005}. The corresponding optical conductivity
\pref{Re.sigma} calculated on the base of the full bubble
\pref{pigi-full} reads
\begin{equation}
\lb{cond-general}
        \sigma(\omega) = \frac{e^2}{\omega} \mbox{Im}
        \int d\bk\, \int dx_1 dx_2
        \frac{n_F (x_1)- n_F (x_2)}
        {\omega - x_1 +x_2+i0}
        \mbox{Tr}[\gamma_{i}(\bk)
         A(\bk,x_1)
        \Gamma_{i}(\bk, \omega)
         A(\bk,x_2)],
\end{equation}
where the integration is over RBZ and $2 \pi i A(\omega,\bk) =
G_{A}(\omega,\bk) - G_R(\omega,\bk)$ is the spectral function
expressed via the difference of advanced and retarded GF. Note that
to regularize the electrical conductivity in the limit $\omega\to 0$ one
has to consider the scattering by impurities which is included in
the simplest form by introducing a finite quasiparticle lifetime,
$\tau_{qp}$. The expression for $\sigma(\omega)$ becomes rather
complicated when the solution of \pref{vertex-eq} is substituted
there, but in the limits $\omega \to 0$ and $\omega \tau_{qp} \gg 1$
it can be significantly simplified. These two limits are considered
below.

\subsection{Conductivity for $\omega \to 0$ and $\omega \tau_{qp} \gg 1$}

According to the results of Refs. \cite{Benfatto:2005:PRB,Aristov:2005}, in
the limit $\omega \to 0$ the conductivity $\sigma(\omega)$ reduces to the
``naive'' expression
\begin{equation}
     \label{conducNaive}
        \sigma(\omega)_{naive} =  \frac{e^2}{\omega} \mbox{Im}
        \int d\bk\, \int dx_1 dx_2
        \frac{n_F (x_1)- n_F (x_2)}
        {\omega - x_1 +x_2+i0}
        \mbox{Tr}[\Gamma_{i}(\bk, 0)
        A(\bk,x_1)
        \Gamma_{i}(\bk, 0)
        A(\bk,x_2)],
\end{equation}
with the static vertex $\Gamma_i(\bk,0)$. The dc limit of
$\sigma(\omega)_{naive}$ was obtained also in
Ref.~\cite{Sharapov:2003:PRB}, where the link between the present
problem and the approach developed by Langer \cite{Langer:1962a:PR}
was pointed out.

Comparing  the LHS of Eqs.~\pref{WI-general} and \pref{Ward-DDW} which
enter, respectively, in  the current-current correlation function
for DDW model \pref{pigi-DDW} and in Eq.~\pref{conducNaive}, one can
notice that $\sigma(\omega)_{naive}$ coincides with the {\em exact
\/} $\sigma(\omega)$ derived from Eq.~\pref{pigi-DDW} for the
reduced DDW model. This coincidence was the main motivation for the
work of Ref.~\cite{Benfatto:2005:PRB}, where the conductivity
$\sigma(\omega)_{naive}$, which is just an {\em approximation\/} for
the full model \pref{Hamiltonian} with $H_{int}$ given by
Eqs.~\pref{Heisenberg},  was extrapolated to higher $\omega$ and
studied in detail.

The analysis of Ref.~\cite{Aristov:2005} clarifies that
Eq.~\pref{conducNaive} does not remain valid for higher $\omega$.
Considering the general shape of the optical conductivity depicted
in Fig.~\ref{fig:conductivity}, one finds that: (i) as far as the
Drude peak is concerned, for $\omega \tau_{qp} \gg 1$ the current
vertex $\Gamma_{i}(\bk,\omega)$ changes from its $\omega=0$ value
Eq.~\pref{WI-general} to its bare value Eq.~\pref{bar-vertex}.
Accordingly, the value of $\sigma(\omega)$ already near the tail of
the Drude peak may well be estimated using Eq.~\pref{cond-general}
with two bare vertices $\gamma_i^0(\bk)$; (ii) as far as intraband
transitions are concerned, the vertex correction enhance $\s(\o)$
only near the edge of the intraband optical absorption. Thus the
expression \pref{conducNaive} overestimates the value of
$\sigma(\omega)$ both in the Drude part, for $\omega \tau_{qp} \gg
1$, and in the region of intraband transitions.  A somewhat
counterintuitive result of Ref.~\cite{Aristov:2005} is that the
crossover from the full vertex $\Gamma_i(\bk,\omega)$ to the bare
vertex $\gamma_{i}^0(\bk)$ occurs at $\omega \sim \tau_{qp}^{-1}$
rather than at the energy scale $D_0$ related to the formation of
the DDW order. It would be interesting to study how this result
depends on the choice of the microscopic $H_{int}$. We note also the
influence of the full vertex $\Gamma_i(\bk,\omega)$ discussed here
is obtained under the assumption that the impurity scattering rate
$1/2\tau_{qp}$ is isotropic. As shown recently in
Ref.~\cite{Gerami:2005}, these results may change drastically if
$1/2\tau_{qp}$ acquires a $\bk$-dependence.

\subsection{Behavior of the spectral weight}

The temperature dependence of the spectral weight $W(T)$ obtained in
Ref.~\cite{Aristov:2005} turns out to be opposite to the behavior of
$W(D,T)$ considered in Sec.~\ref{sec:reduced_model}, viz. in the DDW
state it {\em decreases\/} as the temperature $T$ decreases and the
DDW gap $D(T)$ opens. This decrease follows from the fact that for
$D_0 > \tau_{qp}^{-1}$ the vertex $\Gamma_i(\bk, \omega) \sim
\gamma_i^0(\bk)$, so that the weight of intraband excitations is
smaller than in the reduced model, where it is crucial for producing
an {\em increase\/} of $W^{DDW}(D,T)$ that the vertex
$\gamma_i(\bk)$ \cite{note-reduced} is effective for all frequencies
$\omega$ (see Sec.~\ref{sec:reduced_model}). Observe that this
result should be expected according to the fact that the kinetic
energy increases when the DDW state is formed. As was explained in
detail in Ref.~\cite{Benfatto:2005:PRB}, when vertex corrections are
properly included in the current-current correlation function the
resulting sum rule is given by Eq.~\pref{weight.band} with a
modified occupation number due to the DDW transition, so that $W(T)$
decreases, in analogy to what $W(\Delta,T)$ given by Eq.~\pref{T.SC}
does in the superconducting state.

The results of Ref.~\cite{Aristov:2005} also show that in the
presence of the full vertex the absolute value of the spectral
weight significantly increases with respect to the estimate made
with the bare vertex $\gamma_i^0(\bk)$. Moreover, this increase is
seen already in the normal state above $T_{DDW}$, so that one can
consider that even when $W(T)$ starts to decrease below $T_{DDW}$,
the absolute value of $W(T)$ is still well above the value of the
weight estimated with the bare vertex $\gamma_i^0(\bk)$. This result
should be contrasted with the analysis of
Ref.~\cite{Toschi:2005,Schachinger:2005:PRB}, where instead it was
shown that the interaction {\em decreases} the absolute value of the
normal-state spectral weight with respect to the non-interacting
case, mainly as a consequence of the quasiparticle renormalization
which is instead absent Ref.~\cite{Aristov:2005}.  Thus, further
investigation is required to understand whether this discrepancy is
related simply to the choice of different microscopic models or it
is caused by the approximations used to evaluate the spectral-weight
behavior.

\section{Open questions and relationship to other systems}
\lb{sec:questions}

Our purpose here was not to convince our reader that the DDW state
is the best candidate to explain the pseudogap behavior observed in
HTSC. Neither the reduced nor the full model show full agreement
with existing experimental optical-conductivity data, due in
particular to the lack of a satisfactory description of the shape of
the optical spectra observed experimentally. In particular, no clear
signature of the interband processes characteristic of the DDW state
has been observed in cuprates \cite{review-oc}, even though some
signatures of finite-frequency peaks have been reported recently
\cite{peaks}, which can still be attributed to some kind of charge
ordering \cite{Benfatto:2004:PRB}.  Moreover, while the reduced DDW
model shows at least an interesting feature that agrees with the
experiments, i.e. an increasing spectral weight in the non SC state,
an attempt to generalize it and consider a more microscopic model
destroys this behavior. Certainly more work has to be done to
understand the link between reduced and full models. Nevertheless we
hope that the results discussed here for the restricted
optical-conductivity sum rule in the DDW state may be useful for the
investigation of this sum rule in more sophisticated models leading
to unconventional charge-density wave ordering or circulating
currents (see e.g. \cite{Varma:1997:PRB}). Since the self-energy in
these models is also $\bk$-dependent, to study the transport
properties consistently one should definitely include vertex
corrections. Finally we mention that the results presented here may
be useful not only to cuprates, where they can be only partly
applied, but also to other materials displaying a true $k$-space
modulated CDW, as for example organic conductors \cite{Maki} and
dichalcogenide materials \cite{Vescoli:1998:PRL,Neto:2001:PRL}.

\section{Acknowledgments}
We express our deep gratitude to N.~Andrenacci, H.~Beck, S.~Caprara,
C.~Di~Castro and V.P.~Gusynin for fruitful collaboration on the
topics discussed in this review. We also thank D.~Aristov,
J.P.~Carbotte, J.~Hwang, V.M.~Loktev, M.~Ortolani, A.~Toschi,
D.~Van~der~Marel for helpful discussions and careful reading of the
manuscript. This work was supported by the Natural Science and
Engineering Council of Canada (NSERC) and by the Canadian Institute
for Advanced Research (CIAR).


\begin{thebibliography}{99}
\bibitem{note-vertex} Note that the arguments of the vertex
functions here are written using the conventions of
Ref.~\cite{Aristov:2005} which are different from
Ref.~\cite{Benfatto:2005:PRB}.
\end{thebibliography}
\end{document}